\documentclass[11pt]{article}
\usepackage[a4paper,margin=1.1in]{geometry}
\usepackage{amsmath,amssymb,amsfonts}
\usepackage{graphicx}
\usepackage{booktabs}
\usepackage[round,authoryear]{natbib}
\usepackage{caption}
\usepackage{microtype}
\newcommand{\ket}[1]{\lvert #1 \rangle}

\newcommand{\braket}[2]{\langle #1 \vert #2 \rangle}
\newcommand{\ev}[1]{\langle #1 \rangle}
\usepackage[colorlinks=true,linkcolor=blue,citecolor=blue,urlcolor=blue]{hyperref}
\usepackage{setspace}
\graphicspath{{figs/}}

\title{Quantum Kernels and the Cross-Section of Stock Returns:\\
Anatomy of a Vanishing Advantage\thanks{%
Manuscript prepared for submission to \emph{Quantitative Finance}.
Code and intermediate data for full replication are available from the
author.}}

\author{Junchi Shen}
\date{July 2026}

\begin{document}
\maketitle

\begin{abstract}
\noindent
Do quantum kernels improve cross-sectional stock return prediction? We run
a controlled horse race on the Chinese A-share market in which a quantum
fidelity kernel, a projected quantum kernel, and a classical RBF control
share identical training subsamples, solver, and tuning budgets, so that
only the kernel is exchanged. On the main evaluation---a point-in-time
universe and 170 walk-forward windows (2012--2025)---no quantum advantage
exists: the fidelity kernel is indistinguishable from its RBF control
($\Delta$IC $=+0.005$, $p=0.42$), and a $2\times2$ design crossing kernel
type with training budget (a Nystr\"{o}m extension to the full
$\approx$38{,}000-observation windows) shows quantum kernels matching, but
never beating, equal-budget linear models; after family-wise correction no
pairwise difference among eleven models is significant, with point
estimates favoring penalized linear regressions throughout. We then
document how the opposite conclusion arises: a 60-window evaluation on a
universe screened with full-sample information makes the same quantum
kernel appear dominant on stability criteria and significantly better than
neural baselines. Interaction characteristics from the anomalies
literature help nothing, quantum or classical; a widened bandwidth grid
reveals an interior optimum rather than the near-classical endpoint a
coarse grid suggests; and the geometric difference, while large throughout
($g \gg 1$), does not predict out-of-sample gains ($\rho = -0.20$). We
propose protocol standards---kernel-swap controls, budget-equalized
comparisons, point-in-time universes, and multiplicity-robust
inference---for empirical claims of quantum advantage in finance.
\medskip

\end{abstract}

\section{Introduction}\label{sec:intro}

The machine-learning turn in empirical asset pricing established that
flexible function approximators extract economically meaningful signal
from large sets of firm characteristics \citep{gu2020empirical,
leippold2022machine}. The models of that literature---penalized linear
regressions, boosted trees, shallow networks---differ chiefly in the
feature spaces they implicitly induce. Kernel methods make the choice
explicit: a kernel $\kappa(x,x')$ fully determines the geometry in which
cross-sectional similarity between stocks is measured. Quantum computing
supplies a structurally new kernel family. Encoding a characteristic
vector $x \in \mathbb{R}^d$ into an $n$-qubit state $\ket{\psi(x)}$ maps
the data into a $2^n$-dimensional complex Hilbert space, and the overlap
$\kappa_Q(x,x') = |\braket{\psi(x)}{\psi(x')}|^2$ is a positive
semi-definite kernel whose entangling structure encodes feature
interactions natively \citep{havlicek2019supervised, schuld2019quantum}.

A growing applied literature reports encouraging results for quantum
kernels on financial prediction tasks. Nearly all of it, however, shares
three vulnerabilities: evaluations over short samples, universes
assembled with hindsight, and comparisons against baselines---typically
small neural networks---that are weak in exactly the small-data regimes
where quantum kernels are deployed. Meanwhile the theoretical QML
literature has grown markedly more cautious: expressive quantum
embeddings concentrate exponentially unless inputs are rescaled by a
bandwidth \citep{shaydulin2022importance, canatar2023bandwidth,
thanasilp2024exponential}; the geometric difference between quantum and
classical Gram matrices is necessary but not sufficient for advantage
\citep{huang2021power}; and on generic classical data, tuned fidelity
kernels rarely beat classical kernels \citep{slattery2023numerical,
kubler2021inductive}. What has been missing is a financial study built
to withstand its own destruction: controls that isolate the kernel,
universes buildable in real time, and enough statistical power to
resolve the effect sizes at stake.

This paper supplies that study, and its arc is the result. We predict
20-day-ahead returns for Chinese A-shares from a rotating top-8 subset of
up to 31 firm characteristics, embedded in an entangling feature map and
regressed by quantum kernel ridge regression (QKRR), under a walk-forward
protocol whose defining discipline is a \emph{kernel-swap control}: the
quantum fidelity kernel, a projected quantum kernel
\citep{huang2021power}, and a classical RBF kernel share identical
training subsamples ($N = 1{,}536$ per window), an identical ridge
solver, and identical per-window hyperparameter budgets, so that
performance differences are attributable to kernel geometry alone.

Our main evaluation---27 characteristics rebuilt from raw exchange and
financial-statement data with announcement-date alignment, a
point-in-time universe (trailing coverage and market-capitalization
screens only), and 170 out-of-sample windows spanning 2012--2025---finds
no quantum advantage of any kind. The fidelity QKRR ties its RBF control
($\Delta\text{IC} = +0.005$, $p = 0.42$), ties the three-layer network of
\citet{gu2020empirical} ($p = 0.90$), and trails plain ridge regression in
point estimate ($\Delta\text{IC} = -0.022$, nominal $p = 0.04$)---a
deficit that a budget-equalized $2\times2$ design reveals to be an
artifact of comparing a 1{,}536-observation kernel machine against a
37{,}800-observation regression: at matched budgets, small or large, no
quantum--classical difference is statistically distinguishable, and no
pairwise difference in the study survives family-wise correction. Penalized
linear models---with or without explicit pairwise interactions in the
spirit of \citet{kozak2020shrinking}---occupy the top five places on
every metric, in every five-year subperiod.

The paper's contribution lies in what comes next: an anatomy of how the
opposite conclusion is produced. On a 60-window evaluation restricted to
2021--2025, with a universe screened by \emph{full-sample} price
coverage---a criterion unimplementable in real time---the very same
quantum kernel appears to be the most stable predictor in a sixteen-model
field: highest ICIR (0.49), highest IC t-statistic (3.78), highest hit
rate (80\%), smallest portfolio drawdown ($-27\%$), all while training on
4\% of the classical benchmarks' observations, and with statistically
significant wins over neural baselines. Every element of that apparent
edge dissolves under the clean protocol; most tellingly, merely
rebuilding the \emph{same period's} universe point-in-time halves the
quantum kernel's IC---though ridge falls nearly as much, and a formal
difference-in-differences test cannot attribute the flattery to the
universe alone. That, we argue, is itself the finding: with sixty
windows one cannot even diagnose which ingredient of one's own design
produced the artifact.

Two further experiments trace the mechanism to its roots. First, we
compile twelve pairwise interaction characteristics with published
evidence---value$\times$momentum \citep{asness1997interaction},
value$\times$profitability \citep{piotroski2000value}, gross
profitability \citep{novymarx2013other}, lottery-demand$\times$beta
\citep{bali2017lottery}, and others---and find they improve nothing:
not ridge, not boosted trees, and not the quantum kernel, whose rotation
pool they actively damage ($\Delta\text{IC} = -0.031$, $p = 0.03$) by
displacing stronger primitive signals from scarce qubits that the
entangling layer would otherwise combine itself. Second, bandwidth diagnostics on a widened eight-point grid
locate the tuned optimum at moderate bandwidths, with performance
collapsing in the exponential-concentration regime---the spectrum that
generalizes is dominated by low-order, classically accessible
terms---and the geometric difference $g$ of \citet{huang2021power},
though far above unity in every window, is uncorrelated (indeed slightly
negatively correlated, $\rho = -0.20$) with realized out-of-sample
gains.

We draw two lessons. Substantively, the equity cross-section---heavily
screened, low signal-to-noise, approximately linear in standardized
characteristics---is a hostile arena for quantum kernels, exactly as the
cautious strand of QML theory predicts for generic classical data.
Methodologically, statistical ``quantum advantage'' in finance is easy
to manufacture unintentionally; we propose that future claims meet three
protocol standards: (i) kernel-swap controls that isolate the kernel
from the pipeline, (ii) point-in-time universes, and (iii) window counts
sufficient to resolve the claimed effect size.

Section~\ref{sec:related} reviews the literature.
Section~\ref{sec:data} describes characteristics, universes, and the
interaction catalog. Section~\ref{sec:method} specifies models and
protocol. Section~\ref{sec:main} presents the main (clean-sample)
results, Section~\ref{sec:anatomy} the anatomy of the spurious
advantage, Section~\ref{sec:interactions} the interaction experiments,
and Section~\ref{sec:mechanism} the bandwidth and geometry mechanism.
Section~\ref{sec:discussion} discusses limitations and
Section~\ref{sec:conclusion} concludes.

\section{Related literature}\label{sec:related}

\paragraph{Quantum kernel methods.}
\citet{havlicek2019supervised} and \citet{schuld2019quantum} established
data-encoding circuits as feature maps whose inner products quantum
hardware estimates natively; \citet{schuld2021supervised} showed
variational quantum models reduce to kernel machines.
\citet{huang2021power} derived generalization bounds separating quantum
from classical learners, introduced the projected quantum kernel, and
proposed the geometric difference as a data-dependent screen. The
bandwidth literature \citep{shaydulin2022importance, canatar2023bandwidth}
identified input rescaling as the first-order design choice, and
\citet{thanasilp2024exponential} formalized exponential concentration.
\citet{kubler2021inductive} showed quantum kernels help only when the
target function projects onto their top eigenspaces, and
\citet{slattery2023numerical} reported numerical evidence against
fidelity-kernel advantage on classical data. Our results are the
financial-market counterpart of that caution, obtained under controls
the applied finance literature has generally lacked.

\paragraph{Quantum computing in finance.}
Surveys span portfolio optimization, pricing, and risk
\citep{orus2019quantum, egger2020quantum, herman2023quantum}. Empirical
QML-finance studies concentrate on classification with small, static
datasets; walk-forward cross-sectional regression against a realistic
factor zoo, with the controls above, is to our knowledge new.

\paragraph{Machine learning and the cross-section.}
\citet{gu2020empirical} benchmark the model classes we deploy;
\citet{leippold2022machine} replicate on China, where retail dominance
strengthens technical characteristics---consistent with our rotation
logs, in which lottery MAX \citep{bali2011maxing} and short-term
reversal are the most frequently selected features.
\citet{liu2019size} document China-specific factor structure;
\citet{kozak2020shrinking} and \citet{freyberger2020dissecting} supply
the explicit-interaction and nonparametric-additive benchmarks that
bracket our interaction experiments.

\section{Data, universes, and the interaction catalog}\label{sec:data}

\subsection{Characteristics}

All data are daily and cover January 2010 through March 2026. From
exchange snapshots (price, valuation ratios, market capitalization,
turnover) and quarterly financial statements we construct the firm
characteristics in Table~\ref{tab:factors}, grouped by economic
category. Market-based characteristics are point-in-time by
construction. Fundamental characteristics are merged \emph{as of their
announcement dates}: single-quarter income and cash-flow items are
aggregated to trailing-twelve-month sums only when four consecutive
quarters span at most 380 days, balance-sheet items enter at their
latest disclosed level, year-over-year growth requires a 330--400-day
span, and every value is forward-filled for at most 250 trading days.
Each characteristic is winsorized at $\pm3\sigma$ and standardized
cross-sectionally each day over the full market before any universe
restriction. The main study uses the 27 characteristics reconstructible
from raw data back to 2010; the short-sample study of
Section~\ref{sec:anatomy} additionally uses four characteristics
available only from 2020 (RSRS, analyst revisions, and variants),
for 31 in total.

\begin{table}[t]
\centering
\caption{Firm characteristics by economic category (main-study set).}
\label{tab:factors}
\small
\begin{tabular}{p{3.1cm}p{10.6cm}}
\toprule
Category & Characteristics \\
\midrule
Value (4) & book-to-price, sales-to-price, earnings yield, dividend yield \\
Profitability \& quality (9) & ROA, ROE, gross margin, net margin, asset
turnover, CFO-to-net-income, percent accruals, receivables-to-profit,
working-capital-to-sales \\
Growth \& surprises (6) & sales growth, earnings growth, asset growth,
$\Delta$gross margin, standardized earnings surprise, earnings
variability \\
Risk \& leverage (3) & market beta, debt-to-asset, current ratio \\
Technical (5) & 12--1 momentum, 1-month reversal, Bollinger z-score,
PPO, lottery MAX \citep{bali2011maxing} \\
\bottomrule
\end{tabular}
\end{table}

\subsection{Two universes: point-in-time versus static screen}
\label{sec:universes}

The distinction between our two universe constructions carries the
paper's methodological weight.

\emph{Point-in-time universe (main study).} At each trading day $t$,
eligible stocks are those whose trailing 252-day price coverage exceeds
90\%; the investable pool is the top 450 by trailing 20-day mean float
market capitalization. No information beyond $t$ enters. The panel
contains 1.66 million pool stock-days over 2011--2025.

\emph{Static-screen universe (diagnostic study).} The pool is fixed
over the whole evaluation as the 450 stocks with the highest price
coverage across the \emph{full} 2020--2025 sample. This mimics a
common construction in applied studies---and it conditions on survival
through the sample's end, a screen unimplementable in real time. We
retain it deliberately, as the treatment arm of a natural experiment in
research design.

The prediction target is throughout the 20-trading-day forward return,
with prices extended beyond each study's end so no training label is
truncated.

\subsection{A literature-guided interaction catalog}\label{sec:catalog}

Pure pairwise interactions $z_a \times z_b$ of standardized, weakly
correlated characteristics have near-zero univariate correlation with
returns and are therefore invisible both to linear models and to any
univariate-IC screen---while being precisely what an entangling feature
map is built to encode. To test whether such signals exist and whether
the quantum kernel can monetize them, we compiled from a systematic
literature review the twelve interactions in Table~\ref{tab:catalog},
each constructed as the daily cross-sectional product of its
components' z-scores, re-winsorized and re-standardized. Signs are not
imposed from the (largely U.S.-based) sources; the window-level IC
screen orients interactions exactly as it does base characteristics.
Pooled-sample diagnostics confirm the features carry information
distinct from their components: momentum$\times$surprise attains a
pooled IC of $+0.022$ against $-0.029$ and $+0.007$ for its components,
and lottery$\times$reversal $-0.021$ against $+0.08$ for both of its
components.

\begin{table}[t]
\centering
\caption{Interaction catalog: twelve pairwise interactions with
published evidence.}
\label{tab:catalog}
\small
\begin{tabular}{llp{5.6cm}}
\toprule
Interaction & Components & Source \\
\midrule
Value $\times$ momentum & bp $\times$ mom & \citet{asness1997interaction}; \citet{asness2013value} \\
Value $\times$ profitability & bp $\times$ roa & \citet{piotroski2000value}; \citet{fama2015five} \\
Value $\times$ gross margin & bp $\times$ gm & \citet{novymarx2013other} \\
Gross profitability (GP/A) & gm $\times$ turnover & \citet{novymarx2013other} \\
Beta $\times$ lottery demand & beta $\times$ max & \citet{bali2017lottery} \\
Momentum $\times$ revisions & mom $\times$ rev & \citet{chan1996momentum} \\
Momentum $\times$ surprise & mom $\times$ sue & \citet{chan1996momentum} \\
Lottery $\times$ reversal & max $\times$ strev & \citet{nartea2017extreme}; \citet{bali2011maxing} \\
Momentum $\times$ uncertainty & mom $\times$ epsvol & \citet{zhang2006information} \\
Value $\times$ sales growth & bp $\times$ sg & \citet{lakonishok1994contrarian} \\
Surprise $\times$ uncertainty & sue $\times$ epsvol & \citet{francis2007information} \\
Value $\times$ cash-flow quality & bp $\times$ cfo/ni & \citet{piotroski2000value}; \citet{sloan1996accruals} \\
\bottomrule
\end{tabular}
\end{table}

\section{Models and protocol}\label{sec:method}

\subsection{Walk-forward protocol with factor rotation}

At each rebalancing date $t$ (every 20 trading days) the training window
comprises the trailing 252 trading days ending 20 days before $t$, so
every label is realized; cross-sections are sampled every third day.
For each characteristic we compute the window's average cross-sectional
rank IC. The \emph{active set} keeps characteristics with
$|\overline{\text{IC}}| \ge 0.015$ (minimum six), sign-corrected; the
\emph{top-8 set} keeps the eight largest $|\overline{\text{IC}}|$
characteristics and is shared verbatim by all top-8 models, classical
and quantum. Chinese factor premia rotate strongly across regimes, and
the rotation is visible in the logs: EBITDA/EV's sign flips negative
after September 2024, while lottery MAX enters the top-8 in 56 of 60
short-study windows. The main study yields 170 rebalancing dates
(2012--2025); the diagnostic study yields 60 (2021--2025).

\subsection{Quantum feature map and kernels}

Let $x \in \mathbb{R}^8$ collect the sign-corrected top-8 z-scores
clipped to $[-3,3]$, and let $\tilde{x} = \lambda x$ for bandwidth
$\lambda$. The encoding circuit applies $R = 2$ repetitions of an
IQP-style layer on $n = 8$ qubits,
\begin{equation}
\ket{\psi(x)} \;=\; \prod_{r=1}^{R}
\Bigg[
\prod_{q=1}^{n} e^{-i \tilde{x}_q \tilde{x}_{q+1} Z_q Z_{q+1}}
\;
\prod_{q=1}^{n} e^{-i \tilde{x}_q Z_q} H_q
\Bigg] \ket{0}^{\otimes n},
\label{eq:featuremap}
\end{equation}
with ring entanglement (indices mod $n$). Single-qubit phases encode
factor tilts; $ZZ$ phases encode pairwise interactions. The
\emph{fidelity kernel} is $\kappa_Q(x,x') =
|\braket{\psi(x)}{\psi(x')}|^2$. The \emph{projected quantum kernel}
\citep{huang2021power} collects each qubit's Bloch vector
$\phi_q(x) = (\ev{X_q},\ev{Y_q},\ev{Z_q})$ into
$\phi(x) \in \mathbb{R}^{24}$ and sets
$\kappa_P(x,x') = \exp(-\gamma\|\phi(x)-\phi(x')\|^2)$. The bandwidth
governs spectral decay---large $\lambda$ concentrates the kernel toward
the identity, small $\lambda$ toward a constant
\citep{shaydulin2022importance, canatar2023bandwidth}---and is tuned per
window on the grid $\{0.05, 0.1, 0.2, 0.4\}$. All states are computed by
exact statevector simulation (PennyLane, batched
\citealp{bergholm2018pennylane}); Gram matrices follow from one
inner-product matrix multiplication.

\subsection{QKRR and the kernel-swap control}

Given training Gram matrix $K$ and 20-day forward returns $y$, kernel
ridge regression solves $\hat{\alpha} = (K + \alpha I)^{-1} y$ with
$\alpha \in \{10^{-3},10^{-2},10^{-1},1\}$. Because exact KRR is
$O(N^3)$, each window's training set is subsampled to $N = 1{,}536$
observations stratified by date \citep[Nystr\"{o}m extensions would
restore linear scaling;][]{williams2001using}. The \emph{kernel-swap
control} applies the identical subsample, solver, and search budget to
three kernels---fidelity (\textsc{qkrr-fid}), projected
(\textsc{qkrr-pqk}), and classical RBF on the same top-8 inputs
(\textsc{krr-rbf})---so that any performance difference isolates the
kernel geometry.

\subsection{Classical benchmarks}

On the full active set (and separately on the top-8 set), trained on all
$\approx 37{,}800$ window observations: ridge regression
\citep[$\alpha=10$;][]{hoerl1970ridge}; XGBoost (150 trees, depth 3,
learning rate 0.05) \citep{chen2016xgboost}; a two-hidden-layer MLP
(64--32); and the three-hidden-layer network (32--16--8, three-seed
ensemble) that is the strongest model in \citet{gu2020empirical} and
\citet{leippold2022machine} (\textsc{nn3}). Interaction-aware
benchmarks: \textsc{poly2ridge}, ridge on all pairwise products of the
top-8 characteristics in the spirit of \citet{kozak2020shrinking};
\textsc{ridge-int}, ridge on the active set plus catalog interactions
passing the same IC screen; \textsc{xgb-x}, XGBoost on that extended
set; and \textsc{qkrr-x}, the fidelity QKRR whose top-8 rotation pool
includes the catalog interactions. A rank-average \textsc{hybrid} of
\textsc{qkrr-fid} and ridge completes the field.

\subsection{Hyperparameters, diagnostics, and evaluation}

Within each window the kernel subsample splits temporally 80/20; the
selection criterion for $(\lambda, \alpha, \gamma)$ is validation rank
IC---the quantity a cross-sectional strategy monetizes. As a diagnostic
we compute per window, on a 400-point subset, the regularized geometric
difference \citep{huang2021power}
\begin{equation}
g\big(K_c \,\|\, K_q\big) =
\sqrt{\Big\| \sqrt{K_q}\,\big(K_c + \lambda_g N I\big)^{-1}
\sqrt{K_q} \Big\|_\infty}, \qquad \lambda_g = 10^{-6}.
\label{eq:geodiff}
\end{equation}
Evaluation is identical for all models: out-of-sample cross-sectional
rank IC at each rebalancing date (summarized by mean, ICIR,
$t$-statistic, hit rate; differences tested by paired $t$ and Wilcoxon);
and a long-only top-30\% equal-weighted portfolio rebalanced every 20
days at 23\,bp one-way cost.

\section{Main results: no advantage on the clean sample}
\label{sec:main}

\begin{table}[t]
\centering
\caption{Main evaluation: point-in-time universe, 170 out-of-sample
windows (2012--2025).}
\label{tab:main}
\small
\begin{tabular}{lccccc}
\toprule
Model & Mean IC & ICIR & $t$-stat & Hit rate & Sharpe \\
\midrule
Poly(2) ridge (top-8)          & \textbf{0.0499} & \textbf{0.272} & \textbf{3.55} & 0.629 & 0.272 \\
Ridge (top-8)                  & 0.0494 & 0.247 & 3.21 & \textbf{0.671} & \textbf{0.306} \\
Ridge (full set)               & 0.0477 & 0.248 & 3.23 & 0.647 & 0.272 \\
Hybrid (QKRR $+$ ridge)        & 0.0453 & 0.250 & 3.26 & 0.635 & 0.283 \\
Ridge $+$ catalog interactions & 0.0451 & 0.243 & 3.17 & 0.653 & 0.242 \\
XGBoost (top-8)                & 0.0385 & 0.247 & 3.22 & 0.606 & 0.219 \\
XGBoost (full set)             & 0.0299 & 0.197 & 2.57 & 0.594 & 0.203 \\
QKRR fidelity                  & 0.0254 & 0.171 & 2.23 & 0.582 & 0.134 \\
NN3 \citep{gu2020empirical}    & 0.0243 & 0.202 & 2.64 & 0.529 & 0.137 \\
KRR-RBF control                & 0.0208 & 0.161 & 2.10 & 0.582 & 0.162 \\
QKRR projected                 & 0.0168 & 0.134 & 1.74 & 0.571 & 0.092 \\
\bottomrule
\end{tabular}
\end{table}

Table~\ref{tab:main} contains the paper's central evidence, and its
message is unambiguous. Penalized linear models occupy the top five
places on mean IC; the best of them, the KNS-style pairwise-interaction
expansion, is statistically indistinguishable from its purely linear
counterpart ($\Delta\text{IC} = +0.0004$, $p = 0.94$)---the
cross-section is approximately linear in standardized characteristics,
and what nonlinearity exists is not stably exploitable.

\begin{figure}[t]
\centering
\includegraphics[width=\textwidth]{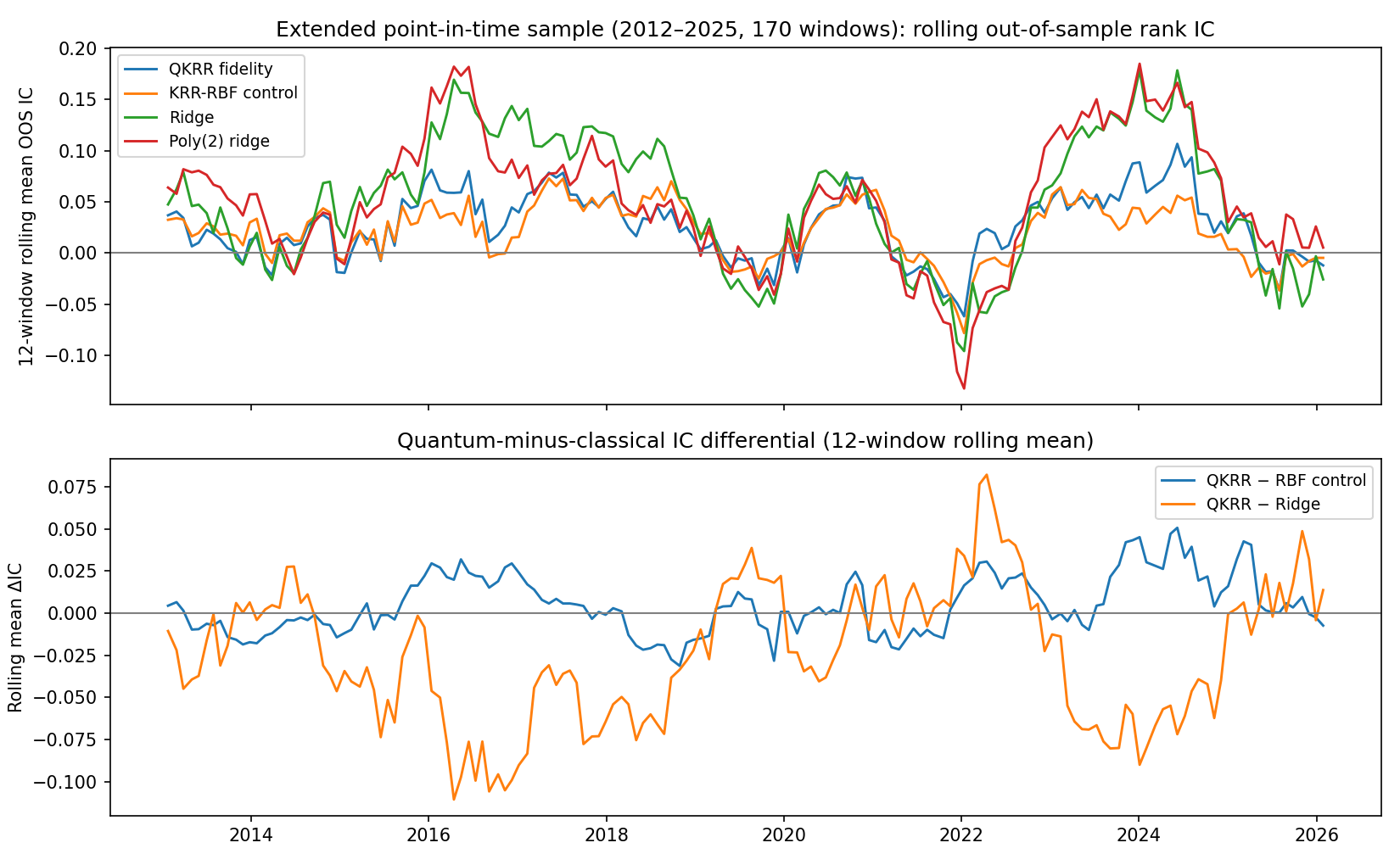}
\caption{Main evaluation, 170 windows. Top: 12-window rolling mean
out-of-sample rank IC. Bottom: rolling quantum-minus-classical IC
differentials.}
\label{fig:main}
\end{figure}

Within the kernel-swap triplet, the fidelity quantum kernel adds nothing
over the classical RBF geometry: $\Delta\text{IC} = +0.005$
($p = 0.42$ paired $t$, $p = 0.66$ Wilcoxon) over 170 windows. Against
models outside the triplet the quantum kernel is either tied or beaten:
$+0.001$ versus \textsc{nn3} ($p = 0.90$), $-0.013$ versus top-8 XGBoost
($p = 0.16$), and $-0.022$ versus plain ridge (nominal $p = 0.043$;
Wilcoxon $p = 0.027$). The last comparison, however, confounds the kernel
with its computational constraint: exact KRR is $O(N^3)$ and trains on
1{,}536 observations while ridge trains on all 37{,}800.
Section~\ref{sec:2x2} disposes of the confound---and of the deficit's
significance. The
quantum--ridge hybrid no longer improves on ridge alone ($p = 0.59$).
The ranking is stable across subperiods: the fidelity QKRR trails ridge
in 2011--2015 (0.027 vs.\ 0.044), 2016--2020 (0.032 vs.\ 0.070), and
2021--2025 (0.020 vs.\ 0.034) alike. Portfolio results agree
(final column of Table~\ref{tab:main}): the quantum portfolio's net
Sharpe of 0.13 sits below the equal-weight benchmark's 0.15, while
top-8 ridge reaches 0.31. Figure~\ref{fig:main} shows the rolling IC
differentials: the quantum-minus-RBF series oscillates around zero
without trend, and the quantum-minus-ridge series is predominantly
negative.

\subsection{Equalizing the training budget: the kernel--budget $2\times2$}
\label{sec:2x2}

Table~\ref{tab:2x2} completes the design by crossing model class with
training budget on the same 170 windows. The small-budget linear control
(\textsc{ridge-sub}) trains on \emph{exactly} the kernel machines'
1{,}536-observation subsamples; the full-budget quantum model
(\textsc{qkrr-nys}) is a Nystr\"{o}m extension \citep{williams2001using}
of the fidelity QKRR to all $\approx$37{,}800 window observations, using
the subsample as landmarks and each window's already-selected bandwidth.

\begin{table}[t]
\centering
\caption{Kernel type $\times$ training budget: mean out-of-sample rank
IC over the same 170 windows. The full-budget quantum entry is a
Nystr\"{o}m extension with the 1{,}536-observation subsample as
landmarks.}
\label{tab:2x2}
\small
\begin{tabular}{lcc}
\toprule
 & $N = 1{,}536$ & $N \approx 37{,}800$ \\
\midrule
Ridge (top-8 inputs) & 0.0355 & 0.0494 \\
QKRR fidelity        & 0.0254 & 0.0438 \\
KRR-RBF control      & 0.0208 & --- \\
\bottomrule
\end{tabular}
\end{table}

The $2\times2$ overturns the one nominally significant comparison in
Table~\ref{tab:main}. At the matched small budget, the quantum kernel
trails the linear model by $-0.010$---insignificant ($p = 0.26$). At the
matched full budget, the Nystr\"{o}m QKRR reaches 0.0438 against ridge's
0.0494---again insignificant ($-0.006$, $p = 0.43$). The apparent
significant deficit to ridge was the sample-budget confound: both model
classes gain comparably from data (quantum $+0.018$, $p = 0.07$; linear
$+0.014$, $p = 0.06$), and once budgets are matched, at either scale,
nothing separates them. This cuts both ways: it removes the significant
quantum \emph{loss}, and it removes the sample-efficiency \emph{selling
point}---the quantum kernel is as data-hungry as everything else, and at
no budget does it beat its classical counterparts.

\subsection{Multiplicity}\label{sec:multiplicity}

The study reports many pairwise comparisons, and at conventional
thresholds some nominal $p$-values in the $0.03$--$0.05$ range will arise
by chance. Holm-adjusting each test family confirms that nothing in the
main evaluation survives correction: the smallest adjusted $p$-value in
the ten-pair family is $0.43$ (fidelity QKRR versus ridge). For a
negative-result paper the correction is welcome news---the core
conclusion of no quantum advantage is multiplicity-robust by
construction---and the same discipline applies symmetrically to the
diagnostic study of Section~\ref{sec:anatomy}, where it proves equally
corrosive to the apparent quantum wins.

\section{Anatomy of a vanishing advantage}\label{sec:anatomy}

Had we stopped at a shorter, more conventional design, this paper would
have reported the opposite result. This section documents that design
and decomposes its flattery of the quantum model.

\subsection{The short-sample study and its apparent quantum edge}

The diagnostic study evaluates the identical protocol on 60 windows
(2021--2025) with the static-screen universe of
Section~\ref{sec:universes} and the full 31-characteristic set.
Table~\ref{tab:short} summarizes. The fidelity QKRR posts a mean IC of
0.0512 with the \emph{best} ICIR (0.488), $t$-statistic (3.78), hit rate
(80\%), and smallest portfolio drawdown ($-27.1\%$) of all sixteen
models---while training on 1{,}536 observations per window against
37{,}800 for the classical benchmarks. It beats the MLP by
$\Delta\text{IC} = +0.040$ ($p = 0.006$) and \textsc{nn3} by $+0.025$
($p = 0.053$), exceeds XGBoost and the RBF control by $+0.015$ each
(insignificant at $p = 0.35$ and $0.17$), and the quantum--ridge hybrid
attains the field's best mean IC (0.0606). Presented alone, these
numbers would read as a success story for quantum kernels---best
stability, sample efficiency, significant wins over deep baselines.

\begin{table}[t]
\centering
\caption{Diagnostic study: static-screen universe, 60 windows
(2021--2025). Selected models.}
\label{tab:short}
\small
\begin{tabular}{lccccc}
\toprule
Model & Mean IC & ICIR & $t$-stat & Hit rate & Max DD \\
\midrule
Hybrid (QKRR $+$ ridge) & \textbf{0.0606} & 0.469 & 3.63 & 0.783 & $-31.3\%$ \\
Ridge (full set)        & 0.0567 & 0.389 & 3.02 & 0.717 & $-33.7\%$ \\
QKRR fidelity           & 0.0512 & \textbf{0.488} & \textbf{3.78} & \textbf{0.800} & $\mathbf{-27.1\%}$ \\
Ridge (top-8)           & 0.0507 & 0.344 & 2.66 & 0.633 & $-33.2\%$ \\
Poly(2) ridge (top-8)   & 0.0470 & 0.354 & 2.74 & 0.617 & $-29.8\%$ \\
XGBoost (full set)      & 0.0366 & 0.320 & 2.48 & 0.700 & $-32.5\%$ \\
KRR-RBF control         & 0.0365 & 0.429 & 3.32 & 0.717 & $-32.8\%$ \\
QKRR projected          & 0.0316 & 0.372 & 2.88 & 0.717 & $-31.8\%$ \\
NN3                     & 0.0258 & 0.328 & 2.54 & 0.617 & $-32.1\%$ \\
MLP (full set)          & 0.0110 & 0.147 & 1.14 & 0.533 & $-33.7\%$ \\
\bottomrule
\end{tabular}
\end{table}

\subsection{Decomposing the illusion}

Three ingredients jointly produce the apparent edge. We examine each,
and---importantly---test whether the decomposition itself can be pinned
down statistically.

\emph{Universe construction.} The static screen conditions on
full-sample data availability---stocks that survived, continuously
traded, and were never delisted through 2025---and its repair moves the
numbers most. Matching the 60 static-universe windows to their nearest
point-in-time counterparts within the \emph{same} 2021--2025 period
($\pm$10 days, 60 pairs), the quantum kernel's mean IC falls from
0.0512 to 0.0184; but ridge falls too, from 0.0567 to 0.0330. A formal
difference-in-differences test of the quantum kernel's \emph{excess}
sensitivity to the universe yields $+0.009$ with $p = 0.71$ (bootstrap
95\% CI $[-0.039, +0.058]$): the clean universe deflates both models,
and the attribution of the quantum kernel's \emph{relative} flattery to
survivorship specifically, while suggested by the point estimates,
cannot be established at window-level power. We flag this honestly---and
note that it sharpens, rather than blunts, the methodological moral:
a 60-window design lacks the power not only to establish an advantage
but even to diagnose the provenance of its own artifacts.

\emph{Insufficient power and multiplicity.} Sixty windows cannot
resolve IC differentials of $\pm 0.015$: the ``stability'' superlatives
(ICIR, drawdown) sat within sampling noise, as their sign reversal
under extension to 170 windows demonstrates. Holm-adjusting the
diagnostic study's own eight-pair test family leaves no comparison
significant (smallest adjusted $p = 0.19$); across both studies, the
only quantum win that survives family-wise correction anywhere is the
comparison against the MLP---the weakest model in the field.

\emph{Weak deep baselines.} That surviving win is telling: the only
(even nominally) significant quantum victories in either study are
against neural networks, the model class most starved at this data
scale. Against properly regularized linear models the quantum kernel
never held a significant advantage in any sample, at any budget.

\section{Interactions do not rescue the quantum kernel}
\label{sec:interactions}

The entangling layer of Eq.~\eqref{eq:featuremap} is an interaction
engine, and Section~\ref{sec:catalog} showed that univariate screens are
structurally blind to pure interactions. A natural conjecture is that
supplying documented interaction signals directly would be worth most
to the quantum model. The data reject the conjecture at every level
(short-sample study; the clean-sample evaluation of
Table~\ref{tab:main} confirms the classical half).

\emph{Interactions do not help classical models.} Ridge with catalog
interactions scores 0.0531 against 0.0567 without; the KNS-style
poly(2) expansion scores 0.0470 against 0.0507 for linear top-8. On the
extended sample the corresponding gaps are $-0.003$ ($p = 0.33$) and
$+0.0004$ ($p = 0.94$). Twenty years of documented interaction
anomalies, constructed faithfully, do not beat their own linear
components out of sample in this market.

\emph{Interactions actively hurt the quantum kernel.} \textsc{qkrr-x},
whose rotation pool admits catalog interactions (they enter the top-8 in
53\% of windows), scores 0.0199 against 0.0512 for the unmodified
\textsc{qkrr-fid} ($\Delta\text{IC} = -0.031$, nominal $p = 0.032$;
the point estimate is large though the comparison, like all others in
its family, does not survive Holm correction). The
mechanism is threefold: the $ZZ$ layer already generates products
internally, making explicit product features redundant; qubit capacity
is scarce, so an admitted interaction displaces a base characteristic
whose stronger, more stable IC was worth more; and product features
inherit noise multiplicatively, degrading the rotation itself. The
design rule for quantum feature maps in finance follows: \emph{spend
qubits on the strongest primitive signals and let entanglement build
the interactions; never hand the circuit pre-computed products.}

\section{Mechanism: bandwidth and geometry}\label{sec:mechanism}

Two diagnostics, tracked window by window, explain why no genuinely
quantum structure ever earned out-of-sample rents.

\emph{Generalization confines the kernel to its low-order regime.} On
the production grid $\{0.05, 0.1, 0.2, 0.4\}$, validation selects the
smallest bandwidth in 24 of 60 short-study windows, and the quantum
kernel's edge over the RBF control concentrates entirely there
($\Delta\text{IC} = +0.043$ in those windows; $\approx 0$ or negative
elsewhere). A referee might suspect---correctly---a grid-endpoint
effect, so we re-ran the bandwidth selection on a widened eight-point
grid spanning $\lambda \in [0.01, 1.6]$. The pile-up at the smallest
point disappears: selections concentrate at interior values
($\lambda = 0.2$ in 16 and $0.4$ in 13 of 60 windows), and the mean
validation IC is single-peaked at $\lambda = 0.2$ (0.163), decaying
mildly toward small bandwidths (0.123 at $\lambda = 0.01$) and
collapsing in the concentration regime (0.104 at $\lambda = 0.8$,
0.077 at $1.6$) \citep{thanasilp2024exponential}. The robust statement
is therefore not that tuning drives the kernel to a classical
\emph{endpoint}, but that the bandwidths which generalize are those at
which the kernel's spectrum is dominated by low-order Taylor
terms---polynomial-like structure that classical kernels replicate---
while the genuinely quantum, high-order interference regime is exactly
where generalization dies. Either way, no channel remains through which
quantum structure could earn out-of-sample rents.

\begin{figure}[t]
\centering
\includegraphics[width=\textwidth]{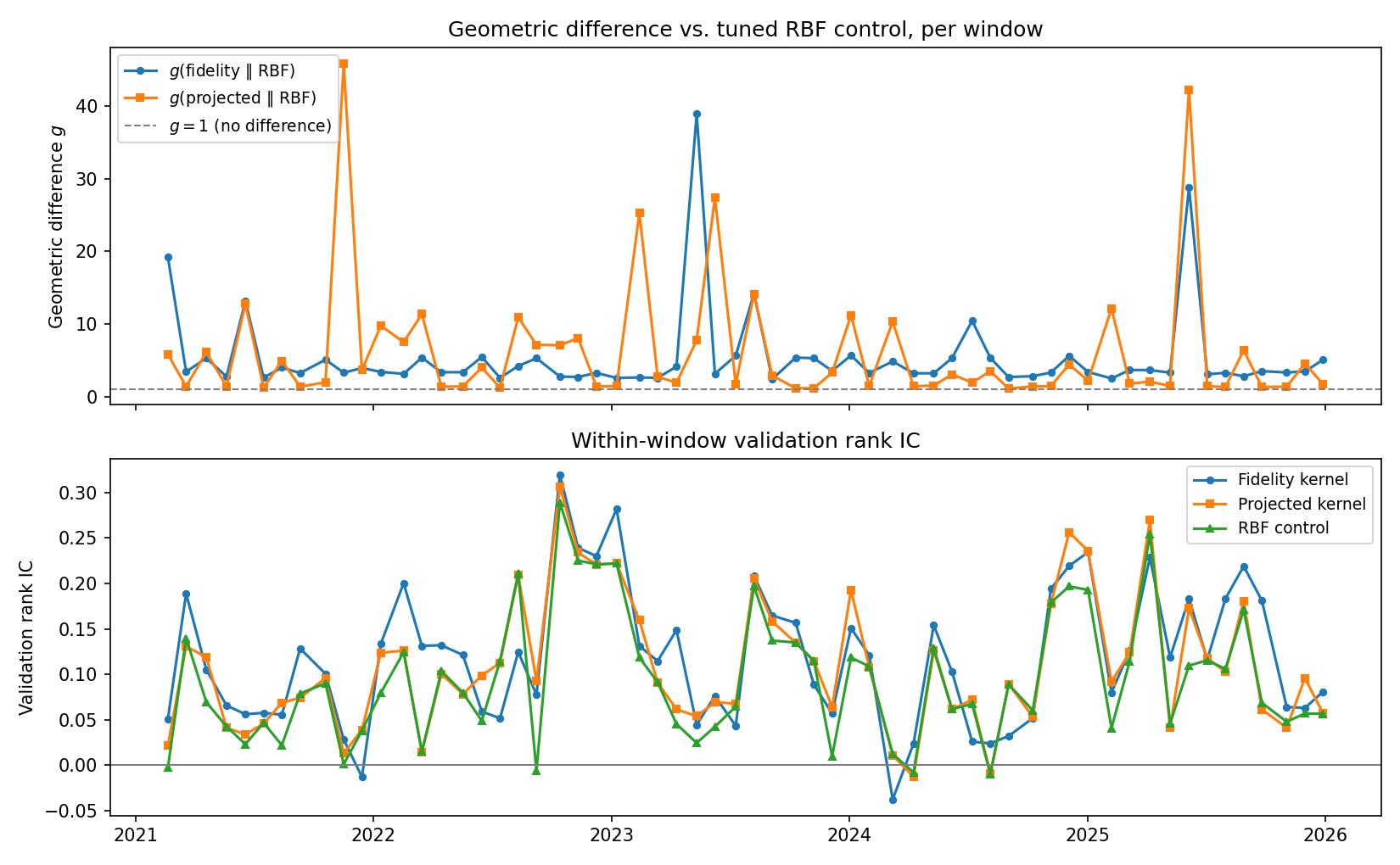}
\caption{Short-sample study, 60 windows. Top: regularized geometric
difference $g$ between quantum kernels and the tuned RBF control.
Bottom: within-window validation rank ICs of the three kernels.}
\label{fig:geo}
\end{figure}

\emph{The geometric difference does not predict gains.} The regularized
$g$ of Eq.~\eqref{eq:geodiff} averages 5.4 (median 3.4, never below
2.4): the quantum geometry is at all times distinct from the tuned RBF
geometry, satisfying the necessary condition of \citet{huang2021power}
throughout. Yet the window-level correlation between $g$ and the
realized quantum-minus-RBF IC differential is $-0.20$. Different
geometry is not better-aligned geometry: the label must live in the
subspace the classical kernel cannot reach, and in this market it does
not \citep{kubler2021inductive}. Figure~\ref{fig:geo} displays both
series.

\section{Discussion and limitations}\label{sec:discussion}

\paragraph{No hardware, and deliberately so.} All kernels are computed
by noiseless statevector simulation; at $n = 8$ qubits the kernel is
classically computable and no computational speedup was ever at stake.
This is the correct arena for the question we ask---whether the kernel
\emph{family} carries statistical value---because hardware noise could
only subtract from the simulated ceiling we measure.

\paragraph{Scope.} One market, one horizon, one feature-map family.
The 20-day horizon and long-only design reflect A-share shorting
constraints; richer encodings (data re-uploading, learned kernel
alignment) might lift the quantum models, though every mechanism we
document---bandwidth confinement, geometry--payoff disconnect,
interaction scarcity---operates upstream of the ansatz choice. Transaction costs
are proportional; neural baselines are deliberately not re-tuned per
window, mirroring standard practice \citep{gu2020empirical}.

\paragraph{Where quantum kernels might still matter in finance.}
Our negative result is conditional on this market's statistical
character: heavy screening, low signal-to-noise, approximate linearity.
The quantum kernel did match full-sample deep networks on 4\% of the
data in both studies---though the $2\times2$ design shows it gains from
additional data just as classical models do, so this is kernel-machine
sample efficiency, not a quantum property. Regimes where usable history
is genuinely short (post-structural-break markets, newly listed assets,
crypto cross-sections) remain a natural test bed, provided the protocol
standards below travel along. Group-structured data with symmetries
matched to covariant kernels \citep{glick2024covariant} and
quantum-generated data are the theoretically favored arenas.

\section{Conclusion}\label{sec:conclusion}

On a point-in-time universe with fourteen years of data and the
statistical power to resolve the effect sizes at stake, quantum kernel
ridge regression offers no advantage for cross-sectional return
prediction in Chinese A-shares: it ties its classical kernel control at
the matched budget, ties equal-budget linear models at full scale under
a Nystr\"{o}m extension, ties the strongest deep-learning benchmark,
and---once training budgets are equalized and inference is corrected
for multiplicity---neither beats nor significantly loses to any
well-specified classical model, while penalized linear
regressions---indifferent to explicit interaction terms---sit atop
every point-estimate ranking. The same pipeline, evaluated the way much
of the applied literature evaluates, would have reported a quantum
success: best stability metrics, apparent sample efficiency,
significant wins over neural baselines. The difference between the two
conclusions is not the model but the protocol---a hindsight-screened
universe, a 60-window sample, and data-hungry baselines suffice to
manufacture a statistical quantum advantage where none exists; and, as
our difference-in-differences test shows, such a design lacks the power
even to diagnose which of its own ingredients did the manufacturing.

We therefore propose that empirical claims of quantum advantage in
finance be held to four standards: a kernel-swap (or equivalent
like-for-like) control that isolates the quantum component;
budget-equalized comparisons that do not confound the kernel with its
computational constraints; universes and features constructible in real
time; and evaluation windows sufficient to resolve the claimed effect
size under family-wise correction. Under those standards, the
productive research frontier is not further horse races on the equity
cross-section, but the search for financial problems whose data are
scarce and whose structure aligns with quantum geometry. Finding one
would be genuine news. Until then, in this market, the cross-section
belongs to ridge regression.

\section*{Data and code availability}
The full pipeline---data construction (both universes), classical
baselines, quantum kernel training, robustness experiments, and
evaluation---is implemented in fourteen Jupyter notebooks with all
intermediate artifacts persisted to disk; quantum simulation uses
PennyLane \citep{bergholm2018pennylane}.
Materials are available from the author on request.

\end{document}